\documentstyle[sprocl]{article}

\def\be{\begin{equation}}
\def\ee{\end{equation}}
\def\bea{\begin{eqnarray}}
\def\eea{\end{eqnarray}}

\def\la{\mathrel{\mathpalette\fun <}}
\def\ga{\mathrel{\mathpalette\fun >}}
\def\fun#1#2{\lower3.6pt\vbox{\baselineskip0pt\lineskip.9pt
  \ialign{$\mathsurround=0pt#1\hfil##\hfil$\crcr#2\crcr\sim\crcr}}}
\def\mpl{{m_{\rm Pl}}}

\begin{document}


\title{MAKING SENSE OF THE NEW COSMOLOGY}

\author{Michael S. TURNER}

\address{Center for Cosmological Physics\\
Departments of Astronomy \&
Astrophysics and of Physics\\
Enrico Fermi Institute, The University of Chicago\\
Chicago, IL~~60637-1433, USA}

\author{}

\address{NASA/Fermilab Astrophysics Center\\
        Fermi National Accelerator Laboratory\\
        Batavia, IL~~60510-0500, USA\\
E-mail: mturner@oddjob.uchicago.edu}


\maketitle\abstracts{Over the past three years we have determined the
basic features of the Universe -- spatially flat; accelerating; comprised
of 1/3 a new form of matter, 2/3 a new form of energy,
with some ordinary matter and a dash of massive neutrinos; and apparently born
from a burst of rapid expansion during which quantum noise
was stretched to astrophysical size seeding cosmic structure.
The New Cosmology greatly extends the highly successful hot
big-bang model.  Now we have to make sense of all this:
What is the dark matter
particle?  What is the nature of the dark energy?  Why this
mixture?  How did the matter -- antimatter asymmetry arise?
What is the underlying cause of inflation (if it indeed
occurred)?}

\section{The New Cosmology}

Cosmology is enjoying the most exciting period of discovery yet.
Over the past three years a new, improved standard cosmology
has been emerging.  It incorporates the highly successful standard
hot big-bang cosmology{\,}\cite{hbb_std} and extends our
understanding of the Universe to times as early as
$10^{-32}\,$sec, when the largest structures in the Universe
were still subatomic quantum fluctuations.

This New Cosmology is characterized by

\begin{itemize}

\item Flat, critical density accelerating Universe

\item Early period of rapid expansion (inflation)

\item Density inhomogeneities produced from quantum fluctuations during inflation

\item Composition:  2/3 dark energy; 1/3 dark matter; 1/200 bright stars

\item Matter content:  $(29\pm 4)$\% cold dark matter; $(4\pm 1)$\% baryons;
$\ga 0.3$\% neutrinos

\item Current temperature:  $T=2.725\pm 0.001\,$K

\item Current age:  $14 \pm 1\,$Gyr

\item Current expansion rate:  $72\pm 7\,{\rm km\,s^{-1}\,Mpc^{-1}}$

\end{itemize}

The New Cosmology is certainly not as well established as the
standard hot big bang.  However, the evidence is mounting.

\subsection{Mounting Evidence:  Recent Results}

The position of the first acoustic peak in the multipole power
spectrum of the anisotropy of the cosmic microwave background
(CMB) radiation provides a powerful means of determining the
global curvature of the Universe.  With the recent DASI observations
of CMB anisotropy on scales of one degree and smaller,
the evidence that the Universe is at most very slightly curved is
quite firm.{\,}\cite{flat}  The curvature radius of the Universe
($\equiv R_{\rm curv}$) and the total energy density parameter
$\Omega_0= \rho_{\rm TOT}/ \rho_{\rm crit}$, are related:
$$ R_{\rm curv} = H_0^{-1}/|\Omega_0-1|^{1/2}$$
The spatial flatness can be expressed as $\Omega_0 = 1.0\pm 0.04$,
or said in words, the curvature radius is at least 50 times greater
than the Hubble radius.

As I will discuss in great detail later, the evidence for accelerated
expansion is also very strong.

The series of acoustic peaks in the CMB multipole power spectrum and
their heights indicate a nearly scale-invariant spectrum
of adiabatic density perturbations with $n=1\pm 0.07$.
Nearly scale-invariant density perturbations and a flat Universe
are two of the three hallmarks of inflation.  Thus, we are beginning
to see the first significant experimental evidence for inflation,
the driving idea in cosmology for the past two decades.

The striking agreement of the BBN determination of the baryon
density from measurements of the primeval deuterium
abundance,{\,}\cite{tytler,baryon_density}
$\Omega_B h^2 = 0.020\pm 0.001$, with those from from recent
CMB anisotropy measurements,{\,}\cite{flat} $\Omega_Bh^2 = 0.022\pm 0.004$,
make a strong case for a small baryon density, as well as the
consistency of the standard cosmology.  (Here, $h=H_0/100\,{\rm km\,
sec^{-1}\,Mpc^{-1}}$.)  There can now be little doubt that baryons
account for but a few percent of the critical density.

Our knowledge of the total matter density is improving, and becoming
less linked to the distribution of light.  This makes determinations
of the matter less sensitive to the uncertain relationship between the
clustering of mass and of light (what astronomers call the bias
factor).{\,}\cite{turner2001}  Both the CMB and clusters of galaxies allow
a determination of the ratio of the total matter density (anything that
clusters -- baryons, neutrinos, cold dark matter) to that in ordinary matter:
$\Omega_M/\Omega_B = 7.2\pm 2.1$ (CMB)
and $=9\pm 1.5$ (clusters).  Not only are these numbers consistent, they
make a very strong case for something beyond quark-based matter.  When
combined with our knowledge of the baryon density, one infers a total
matter density of $\Omega_M = 0.33\pm 0.04$.{\,}\cite{turner2001}

The many successes of the cold dark matter (CDM) scenario  -- from
the sequence of structure formation (galaxies first, clusters
of galaxies and larger objects later) and the structure of the intergalactic
medium, to its ability to reproduce the power spectrum
of inhomogeneity measured today -- makes it
clear that CDM holds much, if not all, of the truth in describing
the formation of structure in the Universe.  This implies that
whatever the dark matter particle is, it moves slowly (i.e., the
bulk of the matter cannot be in the form of hot dark matter such
as neutrinos) and interacts only weakly (e.g., with strength much
less than electromagnetic) with ordinary matter.

The evidence from SuperKamiokande{\,}\cite{superK} for neutrino oscillations
makes a strong case that neutrinos have mass ($\sum_i m_\nu
\ga 0.1\,$eV) and therefore contribute to the mass budget of the
Universe at a level comparable to or greater than that of bright stars.
Particle dark matter has moved from the realm of a hypothesis to
a quantitative question -- how much of each type of particle
dark matter.  Structure
formation in the Universe (especially the existence of small scale
structure) suggests that neutrinos contribute at most 5\% or 10\%
of the critical density.{\,}\cite{cosmo_nulimit}

Even the age of the Universe and the pesky Hubble constant have
been reined in.  The uncertainties in the ages of the oldest globular clusters
have been better identified and quantified, leading to a more precise
age, $t_0 = 13.5\pm 1.5\,$Gyr.{\,}\cite{krauss}  The CMB
can be used to constrain the expansion age, independent of direct
measurements of $H_0$ or the composition of the Universe,
$t_{\rm exp} = 14\pm 0.5\,$Gyr.{\,}\cite{knox}
A host of different techniques are consistent
with the Hubble constant determined by the HST key project,
$H_0 = 72\pm 7\,{\rm km\,s^{-1}\,Mpc^{-1}}$.  Further, the error
budget is now well understood and well quantified.{\,}\cite{H0}
Moreover, the expansion age
derived from this Hubble, which depends upon the composition of the
Universe, is consistent with the previous two ages.

The poster child for precision cosmology continues to be the present
temperature of the CMB.  It was determined by the FIRAS instrument on
COBE to be:  $T_0 = 2.725\pm 0.001\,$K.{\,}\cite{firas}

\subsection{Successes and Consistency Tests}

To sum up; we have determined the basic features of the Universe:
the cosmic matter/energy budget; a self consistent
set of cosmological parameters with realistic errors; and the global
curvature.  Two of the three key predictions of
inflation -- flatness and nearly scale-invariant,
adiabatic density perturbations -- have passed their first significant
tests.  Last but not least, the growing quantity of precision data are
now testing the consistency of the Friedmann-Robertson-Walker framework
and General Relativity itself.

In particular, the equality of the baryon densities determined from
BBN and CMB anisotropy is remarkable.  The first involves nuclear physics
when the Universe was seconds old, while the latter involves gravitational
and classical electrodynamics when the Universe was 400,000 years old.

The entire framework has been tested by the existence of the aforementioned
acoustic peaks in the CMB angular power spectrum.  They reveal large-scale
motions that have remained coherent over hundreds of thousands of years,
through a delicate interplay of gravitational and electromagnetic interactions.

Another test of the basic framework is the accounting of the density of
matter and energy in the Universe.  The CMB measurement of spatial flatness
implies that the matter and energy densities must sum to the critical density.
Measurements of the matter density indicate $\Omega_M = 0.33\pm 0.04$;
and measurements of the acceleration of the Universe from supernovae
indicate the existence of a smooth dark energy component that accounts
for $\Omega_X \sim 0.67$ (depending upon the equation of state).

Finally, while cosmology has in the past been plagued by ``age crises'' -- time
back to the big bang (expansion age) apparently less than the ages of the
oldest objects within the Universe -- today the ages determined by very different
and independent techniques point to a consistent age of 14\,Gyr.

\section{Cosmic Mysteries}

Cosmological measurements and observations over the next decade
will test (and probably refine) the New Cosmology.{\,}\cite{pasp_essay}
If we are fortunate, they
will also help us to make sense of it.  At the moment, there are 
a number of cosmic mysteries, which provide opportunities for surprises and
new insights.  Here I will quickly go through my list, and save the
most intriguing to me -- dark energy -- for its own section at the end.

\subsection{Dark Matter}

By now, the conservative hypothesis is that the dark matter consists of a
new form of matter, with the axion and neutralino as the leading
candidates. That most of the matter in the Universe exists in a new
form of matter -- yet detected in the laboratory -- is a bold and
as of yet untested assertion.

Experiments to directly detect the neutralinos or axions holding our own galaxy
together have now reached sufficient sensitivity to probe the regions
of parameter space preferred by theory.  In addition, the neutralino can
be created by upcoming collider experiments (at the Tevatron or the LHC),
or detected by its annihilation signatures.{\,}\cite{dm_review}

While the CDM scenario is very successful there are some nagging problems.
They may point to a fundamental difficulty or may be explained by messy
astrophysics.{\,}\cite{kosowsky}  The most well known of these problems are
the prediction of cuspy dark matter halos (density profile $\rho_{\rm DM}
\rightarrow 1/r^n$ as $r\rightarrow 0$, with $n\simeq 1 - 1.5$)
and the apparent prediction of too much substructure.  While there are plausible
astrophysical explanations for both problems, they could indicate an
unexpected property of the dark matter particle (e.g., large self interaction
cross section{\,}\cite{pjsdns} or annihilation cross section{\,}\cite{kkt}) or
perhaps even a failure of the particle dark matter paradigm.

I leave for the ``astrophysics to do list'' an accounting of the dark
baryons.  Since $\Omega_B \simeq 0.04$ and $\Omega_* \simeq 0.005$,
the bulk of the baryons are optically dark.  In clusters, the dark
baryons have been identified:  they exists as hot, x-ray emitting
gas.  Elsewhere, the dark baryons have not been identified.  According
to CDM, the bulk of the dark baryons are likely to exist in hot/warm gas
associated with galaxies, but this gas has not been detected.  Since
clusters account for only about 5 percent of the total mass, the bulk
of the dark baryons are still not accounted for.

\subsection{Baryogenesis}

The origin of ordinary, quark-based matter is not yet fully understood.
If the Universe underwent inflation, the dynamical, post-inflation production
of an excess of baryons over antibaryons -- baryogenesis -- is obligatory, since
any pre-inflation baryon asymmetry is diluted away by the enormous entropy
production associated with reheating.

Because we now know that electroweak processes
violate $B+L$ at a very rapid rate at temperatures
above $100\,$GeV or so, baryogenesis is more constrained than when the idea
was introduced more than twenty years ago.  Today there are three
possibilities:  1) produce the baryon asymmetry by GUT-scale physics with $B-L
\not= 0$ (to prevent it being subsequently washed away by $B+L$ violation);
2) produce a lepton asymmetry ($L \not= 0$), which is then transmuted
into the baryon asymmetry by electroweak $B+L$ violation; or
3) produce the baryon asymmetry during
the electroweak phase transition using electroweak $B$ violation.

While none of the three possibilities can be ruled out, the second possibility
looks most promising, and it adds a new twist to the origin of quark-based matter:
We are here because neutrinos have mass.  In the lepton asymmetry first
scenario, Majorana neutrino mass provides the requisite lepton number violation.
The drawback of the first possibility is the necessity of a high reheat temperature
after inflation, $T_{\rm RH} \gg 10^{5}\,$GeV, which is difficult to
achieve in most models of inflation.  The last possibility, while
very attractive because all the input physics might be measurable at
accelerator, requires new sources of $CP$ violation at TeV energies as well
as a strongly first-order electroweak phase transition.

\subsection{Inflation}

There are still many questions to be answered about inflation, including the
most fundamental:  did inflation (or something similar) actually take place!

A good program is in place to test the inflationary framework.  This is done
by testing its three basic predictions:  spatially flat Universe; nearly-scale
invariant, nearly power-law spectrum of Gaussian
adiabatic, density perturbations; and a spectrum of nearly scale-invariant
gravitational waves.

The first two predictions will be probed much more sharply
over the next decade.  The value of
$\Omega_0$ should be determined to much better than 1 percent.
The spectral index $n$ that characterizes the density
perturbations should be measured to percent accuracy.

Generically, inflation predicts $|n-1| \sim
{\cal O}(0.1)$, where $n=1$ corresponds to exact scale invariance.
Likewise, the deviations from an exact power-law predicted by
inflation,{\,}\cite{kosowskymst} $|dn /d\ln k| \sim 10^{-4} - 10^{-2}$
will be tested.  The CMB and the abundance of rare objects such
as clusters of galaxies will allow the prediction of Gaussianity
to be tested.

Inflationary theory has given little guidance as to the amplitude
of the gravitational waves produced during inflation.  If detected,
they are a smokin' gun prediction.  Their amplitude is directly
related to the scale of inflation, $h_{\rm GW} \simeq H_{\rm inflation}/\mpl$.
Together measurements of $n-1$ and $dn/ \ln k$, they can reveal much
about the underlying scalar potential driving inflation.  Measuring their
spectral index -- a most difficult task -- provides a consistency test
of the single scalar-field model of inflation.{\,}\cite{reconstruct}

\subsection{The Dimensionality of Space-time}

Are there additional spatial dimensions beyond the three for which
we have very firm evidence?  I cannot think of a deeper question
in physics today.  If there are new dimensions, they are likely
to be relevant for cosmology, or at least raise new questions in
cosmology (e.g., why are only three dimensions large?  what is going
on in the bulk? and so on).  Further, cosmology may well be
the best means for establishing the existence of extra dimensions.

\subsection{Before Inflation, Other Big-bang Debris, and Surprises}

Only knowing everything there is to know about the Universe
would be worse than knowing all the questions to ask about it.
Without doubt, as our understanding deepens, new questions and
new surprises will spring forth.

The cosmological attraction of inflation
is its ability to make the present state of the Universe insensitive
to its initial state.  However, should we establish inflation as part of
cosmic history, I am certain that cosmologists will begin asking
what happened before inflation.

Progress in cosmology depends upon studying relics.  We have made
much of the handful we have -- the light elements, the baryon asymmetry,
dark matter, and the CMB.  The significance of a new relic cannot
be overstated.  For example, detection of the cosmic sea of neutrinos would
reveal the Universe at 1 second.  Identifying the neutralino
as the dark matter particle and determining its properties at an accelerator
laboratory would open a window on the Universe at $10^{-8}\,$sec.
By comparing its relic abundance as derived from its mass and cross section
with its actual abundance measured in the Universe, one could test
cosmology at the time the neutralino abundance was determined.

And then there are relics that cosmologists have yet to dream of!

\section{Dark Energy:  Seven Lessons}

The dark energy accounts for 2/3 of the stuff in
the Universe and determines its destiny.  That puts it high
on the list of outstanding problems in cosmology.
Its deep connections to fundamental physics -- a new
form of energy with repulsive gravity and possible implications for
the divergences of quantum theory and supersymmetry breaking --
put it very high on the list of outstanding problems
in particle physics.

Dark energy is my term for the causative agent of the current
epoch of accelerated expansion.  According to the second
Friedmann equation,
\begin{equation}
{\ddot R \over R} = -{4\pi G \over 3}\left( \rho + 3 p \right)
\label{eq:acc-eq}
\end{equation}
this stuff must have negative pressure, with magnitude
comparable to its energy density, in order to produce accelerated
expansion [recall $q = -(\ddot R/R)/H^2$; $R$ is the
cosmic scale factor].  Further, since
this mysterious stuff does not show its presence in galaxies
and clusters of galaxies, it must be relatively smoothly distributed.

That being said, dark energy has the following defining properties:
(1) it emits no light; (2) it has large, negative pressure,
$p_X \sim -\rho_X$; and (3) it is approximately
homogeneous (more precisely, does not
cluster significantly with matter on scales at least as large as clusters
of galaxies).  Because its pressure is comparable in magnitude
to its energy density, it is more ``energy-like'' than ``matter-like''
(matter being characterized by $p\ll \rho$).
Dark energy is qualitatively very different from dark matter, and is
certainly not a replacement for it.

It has been said that the sum total of progress in understanding
the acceleration of the Universe is naming the causative agent.
While not too far from the truth, there has been some progress.
I will summarize it below.

\subsection{Two Lines of Evidence for an Accelerating Universe}

Two lines of evidence point to an accelerating Universe.  The first
is the direct evidence based upon measurements of type Ia supernovae
carried out by two groups, the Supernova Cosmology Project{\,}\cite{perlmutter} and
the High-$z$ Supernova Team.{\,}\cite{riess}  These two teams used different
analysis techniques and different samples of high-$z$ supernovae
and came to the same conclusion:  the expansion of the Universe is speeding
up, not slowing down.

The recent serendipitous discovery of a supernovae at $z=1.76$
bolsters the case significantly{\,}\cite{riess2001} and provides the first evidence for
an early epoch of decelerated expansion.{\,}\cite{turner_riess}
SN 1997ff falls right on the accelerating Universe curve
on the magnitude -- redshift diagram,
and is a magnitude brighter than expected in a dusty open Universe
or an open Universe in which type Ia supernovae are systematically
fainter at high-$z$.

The second, independent line of evidence for the accelerating Universe
comes from measurements of the composition of the Universe, which
point to a missing energy component with negative pressure.  The argument
goes like this:  CMB anisotropy measurements indicate that
the Universe is nearly flat, with density parameter, $\Omega_0=1.0\pm 0.04$.
In a flat Universe, the matter density and energy density
must sum to the critical density.  However,
matter only contributes about 1/3
of the critical density, $\Omega_M = 0.33\pm 0.04$.{\,}\cite{turner2001}
(This is based upon measurements of CMB anisotropy,
of bulk flows, and of the baryonic fraction in clusters.)  Thus,
two thirds of the critical density is missing!  Doing the bookkeeping
more precisely, $\Omega_X = 0.67\pm 0.06$.

In order to have escaped detection, this missing energy
must be smoothly distributed.
In order not to interfere with the formation of structure (by
inhibiting the growth of density perturbations),
the energy density in this component must change
more slowly than matter (so that it was subdominant in the past).
For example, if the missing 2/3 of critical density were smoothly
distributed matter ($p=0$), then linear density
perturbations would grow as $R^{1/2}$ rather than as $R$.  The shortfall
in growth since last scattering ($z\simeq 1100$) would be a factor of 30,
far too little growth to produce the structure seen today.

The pressure associated with the missing energy component determines
how it evolves:
\begin{eqnarray}
\rho_X \propto R^{-3(1+w)} \nonumber\\
\rho_X /\rho_M \propto (1+z)^{3w}
\end{eqnarray}
where $w$ is the ratio of the pressure of the missing energy component
to its energy density (here assumed to be constant).
Note, the more negative $w$, the faster the ratio of missing energy
to matter goes to zero in the past.  In order to grow the
structure observed today from the density perturbations indicated by CMB anisotropy
measurements, $w$ must be more negative than about $-{1\over 2}$.{\,}\cite{turnerwhite}

For a flat Universe the deceleration parameter today is
$$q_0 = {1\over 2} + {3\over 2}w\Omega_X \sim {1\over 2} + w$$
Therefore, knowing $w<-{1\over 2}$ implies $q_0<0$ and accelerated expansion.
This independent argument for accelerated expansion and dark energy makes
the supernova case all the more compelling.

\subsection{Gravity Can Be Repulsive in Einstein's Theory, But ...}

In Newton's theory, mass is the source of the gravitational
field and gravity is always attractive.  In General Relativity,
both energy and pressure source the gravitational field.
This fact is reflected in Eq. \ref{eq:acc-eq}.  Sufficiently large
negative pressure leads to repulsive gravity.  

While accelerated expansion can be accommodated within Einstein's theory,
that does not preclude that the ultimate explanation
lies in a fundamental modification of Einstein's theory.
Lacking any good ideas for such a modification, I will discuss
how accelerated expansion fits in the context of General Relativity.
If the explanation for the accelerating
Universe ultimately fits within the Einsteinian framework, 
it will be a stunning new triumph for General Relativity.

\subsection{The Biggest Embarrassment in all of Theoretical Physics}

Einstein introduced the cosmological constant to balance the attractive
gravity of matter.  He quickly discarded the
cosmological constant after the discovery of the expansion
of the Universe.  Whether or not Einstein appreciated that
his theory predicted the possibility of repulsive gravity
is unclear to me.

The advent of quantum field theory made consideration of
the cosmological constant obligatory not optional:  The only
possible covariant form for the energy of the (quantum) vacuum,
$$ T_{\rm VAC}^{\mu\nu} = \rho_{\rm VAC}g^{\mu\nu},$$
is mathematically equivalent to the cosmological constant.
It takes the form for a perfect
fluid with energy density $\rho_{\rm VAC}$ and isotropic
pressure $p_{\rm VAC} = - \rho_{\rm VAC}$ (i.e., $w=-1$)
and is precisely spatially uniform.  Vacuum energy is almost
the {\em perfect} candidate for dark energy.

Here is the rub: the contributions of well-understood physics
(say up to the $100\,$GeV scale) to the quantum-vacuum energy
add up to $10^{55}$ times the present critical density.
(Put another way, if this were so, the Hubble time would
be $10^{-10}\,$sec, and the associated event horizon
would be 3\,cm!)  This is the well known cosmological-constant
problem.{\,}\cite{weinberg,carroll}

While string theory currently offers the best hope for
marrying gravity to quantum mechanics, it has shed precious little
light on the cosmological constant problem, other than to speak 
to its importance. Thomas has
suggested that using the holographic principle to count
the available number of states in our Hubble volume
leads to an upper bound
on the vacuum energy that is comparable to the energy
density in matter + radiation.{\,}\cite{sthomas}  While this
reduces the magnitude of the cosmological-constant
problem very significantly, it does not solve the dark
energy problem:  a vacuum energy that is always comparable
to the matter + radiation energy density would strongly
suppress the growth of structure.

The deSitter space associated with the accelerating Universe may pose
serious problems for the formulation of string theory.{\,}\cite{witten}
Banks and Dine argue that all explanations
for dark energy suggested thus far are incompatible with
perturbative string theory.{\,}\cite{dine_banks}  At the very
least there is high tension between accelerated expansion
and string theory.

The cosmological constant problem leads to a fork in the
dark-energy road:  one path is to wait for theorists to get the
``right answer'' (i.e., $\Omega_X = 2/3$); the other path is to assume that
even quantum nothingness weighs nothing and something else
with negative pressure must be causing the Universe to speed up.
Of course, theorists follow the advice of Yogi Berra:
``When you see a fork in the road, take it.''

\subsection{Parameterizing Dark Energy:  For Now, It's $w$}

Theorists have been very busy suggesting all kinds of interesting
possibilities for the dark energy:  networks of topological defects,
rolling or spinning scalar fields (quintessence and spintessence),
influence of ``the bulk'', and the breakdown
of the Friedmann equations.{\,}\cite{carroll,turner2000}
An intriguing recent paper suggests
dark matter and dark energy are connected through axion
physics.{\,}\cite{barr_seckel}

In the absence of compelling theoretical guidance,
there is a simple way to parameterize dark energy,
by its equation-of-state $w$.{\,}\cite{turnerwhite}

The uniformity of the CMB testifies to the near isotropy
and homogeneity of the Universe.  This implies that the
stress-energy tensor for the Universe must take the perfect
fluid form.{\,}\cite{hbb_std}  Since dark energy dominates
the energy budget, its stress-energy tensor must, to a
good approximation, take the form
\begin{equation}
{T_{X}}^\mu_\nu \approx {\rm diag}[\rho_X,-p_X,-p_X,-p_X]
\end{equation}
where $p_X$ is the isotropic pressure and the desired dark
energy density is
$$\rho_X = 2.7\times 10^{-47}\,{\rm GeV}^4$$
(for $h=0.72$ and $\Omega_X = 0.66$).  This corresponds to
a tiny energy scale, $\rho_X^{1/4} = 2.3\times 10^{-3}\,$eV.

The pressure can be characterized by its ratio to the
energy density (or equation-of-state):
$$w\equiv p_X/\rho_X$$
Note, $w$ need not be constant; e.g., it could be a function of $\rho_X$
or an explicit function of time or redshift.  ($w$ can always
be rewritten as an implicit function of redshift.)

For vacuum energy $w=-1$; for a network of topological defects
$w=-N/3$ where $N$ is the dimensionality of the defects (1 for
strings, 2 for walls, etc.).  For a minimally coupled, rolling scalar field,
\begin{equation}
w = {{1\over 2} \dot\phi^2 - V(\phi ) \over {1\over 2} \dot\phi^2 + V(\phi )}
\end{equation}
which is time dependent and can vary between $-1$ (when potential energy
dominates) and $+1$ (when kinetic energy dominates).  Here $V(\phi )$ is
the potential for the scalar field.

I believe that for the foreseeable future
getting at the dark energy will mean trying
to measure its equation-of-state, $w(t)$.

\subsection{The Universe:  The Lab for Studying Dark Energy}

Dark energy by its very nature is diffuse and a low-energy
phenomenon.  It probably cannot be produced at accelerators;
it isn't found in galaxies or even clusters of galaxies.
The Universe itself is the natural lab -- perhaps the only
lab -- in which to study it.

The primary effect of dark energy on the Universe
is on the expansion rate.  The first Friedmann equation
can be written as
\begin{equation}
H^2(z)/H_0^2  = \Omega_M(1+z)^3 +
    \Omega_X \exp \left[ 3\int_0^z\,[1+w(x)]d\ln (1+x) \right]
\end{equation}
where $\Omega_M$ ($\Omega_X$) is the fraction of critical density
contributed by matter (dark energy) today, flatness is assumed, 
and the dark-energy term follows
from energy conservation, $d(\rho_X R^3) = -p_X
dR^3$.  For constant $w$ the dark energy
term is simply $\Omega_X(1+z)^{3(1+w)}$.  Note, with the assumption
of flatness, $H(z)/H_0$ depends upon only two parameters:  
$\Omega_M$ and $w(z)$.

While $H(z)$ is probably not directly measurable (however
see Loeb{\,}\cite{Loeb}), it does affect two
observable quantities:  the (comoving) distance to an object
at redshift $z$,
$$r(z) = \int_0^z \,{dz\over H(z)},$$
and the growth of (linear) density perturbations, governed by
$$\ddot\delta_k + 2H\dot\delta_k - 4\pi G\rho_M\delta_k = 0,$$
where $\delta_k$ is the Fourier component of comoving
wavenumber $k$ and overdot indicates $d/dt$.

The comoving distance $r(z)$ can be probed by
standard candles (e.g., type Ia supernovae)
through the classic cosmological observable,
luminosity distance $d_L(z) = (1+z)r(z)$.  It can also
be probed by counting objects of a known intrinsic comoving
number density, through the comoving volume
element, $dV/dzd\Omega = r^2(z)/H(z)$.

Both galaxies and clusters of galaxies have been suggested
as objects to count.{\,}\cite{count}  For each,
their comoving number density evolves (in the case of
clusters very significantly).  However, it is believed that
much, if not all, of the evolution can be modelled through
numerical simulations and semi-analytical calculations
in the CDM picture.  In the case of clusters, evolution is
so significant that the number count test probe is affected
by dark energy through both $r(z)$ and the growth of perturbations,
with the latter being the dominant effect.

The various cosmological approaches to ferreting
out the nature of the dark energy have been studied extensively.{\,}\cite{yellowbook}
Based largely upon my work with Dragan Huterer,{\,}\cite{ht}
I summarize what we know about the efficacy of the cosmological
probes of dark energy:

\begin{itemize}

\item Present cosmological observations prefer $w=-1$,
with a 95\% confidence limit $w < -0.6$.{\,}\cite{perlmutteretal}

\item Because dark energy was less important in the past,
$\rho_X/\rho_M \propto (1+z)^{3w}\rightarrow 0$ as $z
\rightarrow \infty$, and the Hubble
flow at low redshift is insensitive to the composition of
the Universe, the most sensitive
redshift interval for probing dark energy is $z=0.2 - 2$.{\,}\cite{ht}

\item The CMB has limited power to probe $w$ (e.g.,
the projected precision for Planck is $\sigma_w = 0.25$)
and no power to probe its time variation.{\,}\cite{ht}

\item A high-quality sample of 2000 SNe distributed from
$z=0.2$ to $z=1.7$ could measure $w$ to a precision $\sigma_w
=0.05$ (assuming an irreducible systematic error of 0.14 mag).
If $\Omega_M$ is known independently to better
than $\sigma_{\Omega_M} = 0.03$, $\sigma_w$ improves by
a factor of three and the rate of change of $w^\prime =
dw/dz$ can be measured to precision $\sigma_{w^\prime} = 0.16$.{\,}\cite{ht}

\item Counts of galaxies and of clusters of galaxies may have
the same potential to probe $w$ as SNe Ia.  The
critical issue is systematics (including the evolution of
the intrinsic comoving number density, and the ability to identify
galaxies or clusters of a fixed mass).{\,}\cite{count}

\item Measuring weak gravitational lensing by large-scale
structure over a field of 1000 square degrees (or more)
could have comparable sensitivity to $w$ as type Ia supernovae.
However, weak gravitational lensing does not appear to be a good
method to probe the time variation of $w$.{\,}\cite{dragan}  The systematics
associated with weak gravitational lensing have not yet been studied
carefully and could limit its potential.

\item Some methods do not look promising in their ability
to probe $w$ because of irreducible systematics (e.g.,
Alcock -- Paczynski test and strong gravitational lensing
of QSOs).  However, both could provide important independent
confirmation of accelerated expansion.

\end{itemize}

\subsection{Why now?:  The Nancy Kerrigan Problem}

A critical constraint on dark energy is that it not interfere
with the formation of structure in the Universe.  This means
that dark energy must have been relatively unimportant in the
past (at least back to the time of last scattering, $z\sim 1100$).
{If} dark energy is characterized by constant $w$, not
interfering with structure formation can be quantified as:
$w\la -{1\over 2}$.{\,}\cite{turnerwhite}  This means
that the dark-energy density evolves more slowly than
$R^{-3/2}$ (compared to $R^{-3}$ for matter) and implies
\begin{eqnarray*}
\rho_X/\rho_M & \rightarrow & 0\qquad{\rm for\ }t\rightarrow 0 \\
\rho_X/\rho_M & \rightarrow & \infty \qquad{\rm for\ }t\rightarrow \infty \\
\end{eqnarray*}

That is, in the past dark energy was unimportant and in the
future it will be dominant!  We just happen to live at the
time when dark matter and dark energy have comparable densities.
In the words of Olympic skater Nancy Kerrigan, ``Why me?  Why now?''

Perhaps this fact is an important clue to unraveling the nature
of the dark energy.  Perhaps not.  I shudder to say this, but
it could be at the root of an anthropic explanation for the size
of the cosmological constant:  The cosmological constant is
as large as it can be and still allow the formation of structures
that can support life.\cite{weinberg2}

\subsection{Dark Energy and Destiny}

Almost everyone is aware of the connection between the shape
of the Universe and its destiny:  positively curved recollapses,
flat; negatively curved expand forever.  The link between
geometry and destiny depends upon a critical assumption: that
matter dominates the energy budget (more precisely, that all components
of matter/energy have equation of state $w> -{1\over 3}$).
Dark energy does not satisfy this condition.

In a Universe with dark energy the connection between geometry
and destiny is severed.{\,}\cite{kraussturner}  A flat Universe
(like ours) can continue expanding exponentially forever
with the number of visible galaxies diminishing to a few
hundred (e.g., if the dark energy is a true cosmological constant);
the expansion can slow to that of a matter-dominated
model (e.g., if the dark energy dissipates and becomes sub
dominant); or, it is even possible for the Universe to recollapse
(e.g., if the dark energy decays revealing a negative cosmological constant).
Because string theory prefers anti-deSitter space, the third
possibility should not be forgotten.

Dark energy then is the key to understanding our destiny.

\subsection{The Challenge}

I believe the really big challenge for the New Cosmology is making sense of dark energy.

Because of its diffuse character, the Universe is likely the lab
where dark energy can best be attacked (though one should not
rule other approaches -- e.g., if the dark energy involves a ultra-light
scalar field, then there should be a new long-range force{\,}\cite{carroll2}).

While type Ia supernovae look particularly promising -- they have
a track record and can, in principle, be used to map out
$r(z)$ -- there are important open issues.
Are they really standardizable candles?  Have they evolved?  Is the
high-redshift population the same as the low-redshift population?

The dark-energy problem is important enough that pursuing complimentary
approaches is both justified and prudent.  Weak-gravitational lensing
shows considerable promise.  While beset by important issues involving number
evolution and the determination of galaxy and cluster masses,{\,}\cite{count}
counting galaxies and clusters of galaxies holds promise for probing
the dark energy.

Two realistic goals for the next decade are determining
$w$ to 5\% and probing the time variation.  Either
has the potential to rule out a cosmological constant:
For example, by measuring a significant time
variation in $w$ or by pinning $w$ at $5\sigma$ away from $-1$.
Such a development would be remarkable and far reaching.

After determining the equation-of-state of the dark energy, the
next step is measuring its clustering properties.
A cosmological constant is spatially constant; a rolling
scalar field clusters slightly on very large scales.{\,}\cite{friemanetal}
Measuring its clustering properties will not be easy,
but it provides an important, new window on dark energy.{\,}\cite{okamoto}

We do live at a special time:  There is still enough light in the Universe
to illuminate the dark energy.

\section{Closing Remarks}

As a New Cosmology emerges, a new set of questions arises.
Assuming the Universe inflated, what is the physics underlying inflation?
What is the dark-matter particle?  How was the baryon asymmetry produced?
Why is the recipe for our Universe so complicated?
What is the nature of the Dark Energy?  Answering these questions will
help us make sense of the New Cosmology as well as revealing
deep connections between fundamental physics and cosmology.
There may even be some big surprises -- time variation of the
constants or a new theory of gravity that eliminates the need
for dark matter and dark energy.

There is an impressive program in place, with telescopes,
accelerators, and laboratory experiments, both in space and
on the ground:  the Sloan Digital Sky Survey; the Hubble
Space Telescope and the Chandra X-ray Observatory; a growing number of
large ground-based telescopes; the Tevatron and B-factories in the
US and Japan; specialized dark-matter
detectors; gravity-wave detectors; a multitude of ground-based and balloon-borne
CMB anisotropy experiments; the MAP satellite (which is already taking
data) and the Planck satellite (to be launched in 2007).  Still to
come are:  the LHC; a host of accelerator and nonaccelerator neutrino-oscillation
and neutrino-mass experiments; the Next Generation Space Telescope;
gravity-wave detectors in space; cluster surveys using x-rays and
the Sunyaev -- Zel'dovich effect.  And in the planning:  dedicated
ground and space based wide-field telescopes to study dark energy,
the next linear collider and on and on.  Any one, or more likely several,
of these experiments will produce major advances in our understanding of
the Universe and the fundamental laws that govern it.

The progress we make over the two decades will determine
how golden our age of cosmology is.

\section*{Acknowledgments}
This work was supported by the DoE (at Chicago and Fermilab)
and by the NASA (at Fermilab by grant NAG 5-7092).

\end{document}